\documentclass[smallcondensed, final]{svjour3}

\usepackage{lipsum}
\usepackage{amsmath, amssymb}
\usepackage{siunitx}
\usepackage{braket}
\usepackage{graphicx}
\usepackage{subfig}
\usepackage{booktabs}
\usepackage{comment}
\usepackage{soul,color}

\newcommand{\ie}        {i.\,e.}
\newcommand{\eg}        {e.\,g.}
\newcommand{\ped}  [1]  {\ensuremath{_\text{#1}}}
\newcommand{\api}  [1]  {\ensuremath{^\text{#1}}}
\newcommand{\nspin}     {\ensuremath{N}}
\newcommand{\tf}        {\ensuremath{t\ped{f}}}
\newcommand{\npop}      {\ensuremath{N\ped{pop}}}
\newcommand{\ngen}      {\ensuremath{N\ped{g}}}
\newcommand{\tournsize} {\ensuremath{N\ped{T}}}
\newcommand{\gap}       {\ensuremath{\Delta}}
\newcommand{\nanc}      {\ensuremath{N\ped{a}}}
\newcommand{\ntot}      {\ensuremath{M}}
\newcommand{\ham}       {H}

\renewcommand{\descriptionlabel}[1]{\hspace{\labelsep}\textbf{#1}}

\title{An evolutionary strategy for finding effective quantum $ 2 $-body Hamiltonians of $ p $-body interacting systems}

\author{G.\,Acampora$^{1}$ \and
        V.\,Cataudella$^{1,2,3}$ \and
        P.\,R.\,Hegde$^{1}$ \and
        P.\,Lucignano$^{2}$ \and 
        G.\,Passarelli$^{1, 2, \dagger}$ \and 
        A.\,Vitiello$^{1}$\thanks{G.\,A. and A.\,V. supervised the research activities related to evolutionary computation.\\
    P.\,R.\,H. and G.\,P. designed and implemented the genetic algorithm, and performed numerical simulations and data analysis.\\
    P.\,L. and V.\,C. supervised the project.\\
    All authors contributed to the preparation of the present manuscript.}
        }

\institute{
    $^1$: Dipartimento di Fisica ``E.\,Pancini'', Universit\`a di Napoli ``Federico II'', Complesso di Monte S.~Angelo, via Cinthia, 80126 Napoli, Italy\\
    $^2$: CNR-SPIN, c/o Complesso di Monte S. Angelo, via Cinthia - 80126 - Napoli, Italy\\
    $^3$: Istituto Nazionale di Fisica Nucleare, Sezione di Napoli, Napoli, 80126, Italy\\
    $^\dagger$Corresponding author: gpassarelli@fisica.unina.it
}

\titlerunning{An evolutionary strategy for finding quantum $ 2 $-body Hamiltonians\dots}
\authorrunning{G.\,Acampora et al.}

\date{June 3, 2019}

\journalname{Quantum Machine Intelligence}

\begin{document}
	
	\maketitle
	\begin{abstract}
	    Embedding $p$-body interacting models onto the $2$-body networks implemented on commercial quantum annealers is a relevant issue.  For highly interacting models, requiring a number of ancilla qubits, that can be sizable and make unfeasible (if not impossible) to simulate such systems. In this manuscript, we propose an alternative to minor embedding, developing a new approximate procedure based on genetic algorithms, allowing to decouple the $p$-body in terms of $2$-body interactions. A set of preliminary numerical experiments demonstrates the feasibility of our approach for the ferromagnetic $p$-spin model, and pave the way towards the application of evolutionary strategies to more complex quantum models. 
	\end{abstract}
	
	\section{Introduction}
	
	 Finding the solution of NP-hard problems requires a  time-to-solution increasing exponentially as a function of the system size~\cite{cook:complexity}. NP-hard tasks can be studied with adiabatic quantum computation~\cite{farhi:quantum-computation,albash:review-aqc}, a heuristic tool for finding the optimal solution to this kind of problems. The D-Wave quantum machines~\cite{harris:d-wave} can perform finite-time adiabatic quantum computation, or quantum annealing. The superconducting architecture of D-Wave processors is built on the Chimera graph~\cite{choi:2008,choi:2011}, a sparsely connected graph that can host $ \nspin \le 2048 $ qubits, with at most $2$-body interactions. However, many interesting problems, including the ferromagnetic $p$-spin model~\cite{derrida:p-spin,gross:p-spin,bapst:p-spin}, can be mapped on fully-connected qubit systems with $p$-body interactions ($p \ge 2$). In order to exploit the available quantum hardware, these problem have to be mapped to effective Hamiltonians~\cite{lucas:np-complete}, containing at most $2$-body interactions. This necessarily implies the introduction of auxiliary degrees of freedom, or ancillae~\cite{biamonte:k-body-decomposition}.
	 The major challenge in this problem is to find the free parameters in the $ 2 $-body Hamiltonian, corresponding to the $ p $-body one, such that the two Hamiltonians share the same spectral properties.
	 
	 In this paper, we show that genetic algorithms can be a powerful tool to optimize the free parameters in the effective $2$-body model, focusing on the ferromagnetic $p$-spin system. Genetic algorithms are stochastic meta-heuristics for finding solutions to optimization problems, inspired by the Darwinian theory of evolution~\cite{goldberg:genetic-algorithms}. The (real) free  parameters to optimize, or genes, are arranged in a chromosome. Many such chromosomes, or individuals, compose a population. The fitness of each individual represents its chances of survival along generations. Choosing an appropriate fitness function is the core of genetic algorithms. As shown with more in-depth in the following Sections, we use the mean square error of the effective spectrum from that of the original Hamiltonian as our fitness function.
	 The idea to apply genetic algorithms is motivated by recent works~\cite{qumi1,ref1}, where this kind of evolutionary algorithms have been successfully exploited to solve optimization problems in quantum computing domain.
	 
	 
	 The ferromagnetic $p$-spin model is equivalent to the Grover search algorithm in the limit of large and odd $p$. However, in this paper we focus on the very simple cases involving small $ p $ ($p = 3$) that can be also analytically addressed.
	 As shown by a set of preliminary experiments involving two simple configurations of ferromagnetic $p$-spin model, the analytic solutions are well-reproduced by the designed genetic algorithm. Moreover, to ensure the validity of our approach, we also simulate a quantum annealing and study the time evolution of the ground state probability for the $p$-spin system and its effective $2$-body counterpart.
	 
	 The rest of the paper is organized as follows. In Section~\ref{sec:problem}, our model Hamiltonian is introduced. In Section~\ref{proposal}, the details about the proposed genetic algorithm including chromosome structure and fitness function are given. Section~\ref{sec:Experimental Results} presents the settings and the results of a set of  preliminary experiments related to the application of the proposed genetic algorithm to two small instances of the $p$-spin model, which we use as benchmarks for the accuracy of our scheme. Conclusions and improvements to be performed in the future are reported in Section~\ref{sec:Conclusions}.

	\section{Problem definition}
	\label{sec:problem}
	
	We consider a system of $ \nspin $ qubits. The two logical states in the computational basis of qubit $ i $ can be equivalently labeled as $ \ket{\sigma_i} $, with $ \sigma_i = \pm 1 $, or $ \ket{x_i} $, with $ x_i = 0, 1 $. The two choices are related by $ \sigma_i = 1 - 2 x_i $. In the following, we will use the $ x_i $ representation, unless stated otherwise. We denote by $ \sigma_i^{k} $, with $ k = x, y, z $, the Pauli matrices acting on the $ i $th qubit. Moreover, we work in natural units and fix $ \hslash = 1 $.
	
	We focus on the ferromagnetic $ p $-spin model~\cite{derrida:p-spin,gross:p-spin}, whose dimensionless classical Hamiltonian reads
	\begin{equation}\label{eq:pspin-hamiltonian-classical}
		E\ped{p} = -\nspin {\left[ \frac{1}{\nspin} \sum_{i = 1}^{\nspin} (1 - 2 x_i) \right]}^p.	
	\end{equation}
	The quantum version of this Hamiltonian reads
	\begin{equation}\label{eq:pspin-hamiltonian}
		\ham\ped{p} = -\nspin {\left( \frac{1}{\nspin} \sum_{i = 1}^{\nspin} \sigma_i^z \right)}^p.
	\end{equation}
	For even $ p $, there are two degenerate ground states due to the $ Z_2 $ symmetry of this model, while for odd $ p $ the ground state is nondegenerate. For $ \nspin \to \infty $ and $ p \to \infty $ ($ p \le \nspin $, $ p $ odd), this model can implement a Grover-like search in adiabatic quantum computation~\cite{grover:search}.
	
	In adiabatic quantum computation, one usually employs the parametric Hamiltonian
	\begin{equation}\label{eq:time-dependent-hamiltonian}
		\ham(s) = A(s) \ham_0 + B(s) \ham\ped{p},
	\end{equation}
	where $ s = t / \tf $ is a dimensionless time and ranges in $ \left[0, 1\right] $, $ \tf $ being the annealing time, and the two functions $ A(s) $ and $ B(s) $ satisfy $ A(0) \gg B(0) $ and $ A(1) \ll B(1) $.
	$ \ham_0 $ is the transverse field Hamiltonian:
	\begin{equation}\label{eq:transverse-field}
		\ham_0 = -\sum_{i = 1}^{\nspin} \sigma_i^x.
	\end{equation}
	The qubit system is prepared in the ground state of $ \ham(0) $ and is evolved by slowly changing the parameter $ s $ towards $ s = 1 $. If the evolution is adiabatic compared to the inverse of the minimal gap $ \gap $ between the instantaneous ground state and the first excited state, the system is found at $ s = 1 $ in the ground state of $ \ham\ped{p} $ with large probability. In this paper, we will use a linear annealing schedule, \ie, $ A(s) = 1 - s $ and $ B(s) = s $.

	Despite the fact that it is analytically solvable, the $ p $-spin model is heavily studied in the context of quantum optimization~\cite{seoane:transverse-interactions,nishimori:inhomogeneous-2,nishimori:reverse-pspin,passarelli:pspin,passarelli:proceeding}, due to its ability to capture the essential feature of NP-hard problems, \ie, the exponentially growing time-to-solution as a function of $ \nspin $. In fact, when $ p > 2 $ and in the thermodynamic limit, the $ p $-spin system undergoes a first-order quantum phase transitions that makes its spectral gap $ \gap $ close exponentially fast as a function of $ \nspin $~\cite{bapst:p-spin}.
	
	However, due to its full-connectivity and the presence of $ p $-body interactions, this model can hardly be embedded in the available quantum hardware. The Chimera graph of latest D-Wave machines only allow to study sparse models with at most $ 2 $-body interactions~\cite{choi:2008,choi:2011}. In order to use D-Wave machines to perform the quantum annealing of the $ p $-spin model, first we have to map its Hamiltonian~\eqref{eq:pspin-hamiltonian} into an effective one, containing only $ 2 $-body interactions, yet still fully connected. Then, using minor embedding~\cite{choi:2008}, this fully-connected effective $ 2 $-body Hamiltonian is mapped onto a sparse model, respecting the topology of the Chimera graph. Both these two steps require the introduction of a certain number $ \nanc $ of ancillary degrees of freedom. In this paper, we will address only the first question and discuss the mapping of the $ p $-spin Hamiltonian with $ p $-body interactions onto the effective fully-connected $ 2 $-body Hamiltonian
	\begin{equation}\label{eq:pspin-effective}
		\ham\ped{p}' = K + \sum_{i = 1}^{\ntot} h_i \sigma_i^z + \sum_{i = 1}^{\ntot} \sum_{j = i + 1}^{\ntot} J_{i, j} \sigma_i^z \sigma_j^z,
	\end{equation} 
	where $ \ntot = \nspin + \nanc $ is the total number of qubits, $ K $ is a constant energy shift, $ h_i $ are local longitudinal fields and $ J_{i, j} $ couples qubits $ i $ and $ j $ ($ j > i $). The corresponding classical effective energy reads
	\begin{equation}\label{eq:pspin-effective-classical}
	    E\ped{p}' = c_0 + \sum_{i = 1}^{\ntot} c_i x_i + \sum_{i = 1}^{\ntot} \sum_{j = i + 1}^{\ntot} d_{i, j} x_i x_j.
	\end{equation}
	Parameters in the two Hamiltonians~\eqref{eq:pspin-effective} and~\eqref{eq:pspin-effective-classical} are related by~\cite{nishimori:perspectives}
	\begin{gather}
	    K = c_0 + \frac{1}{2} \sum_{i = 1}^{\ntot} c_i + \frac{1}{4} \sum_{i = 1}^{\ntot} \sum_{j = i + 1}^{\ntot} d_{i, j}.\\
	    h_i = -\frac{1}{2} c_i - \frac{1}{4} \sum_{i = 1}^{\ntot} d_{i, j} - \frac{1}{4} \sum_{j = 1}^{\ntot} d_{i, j},\\
	    J_{i, j} = \frac{1}{4} d_{i, j}.
	\end{gather}
	All these free parameters are real-valued.
	
	To map the Hamiltonian~\eqref{eq:pspin-hamiltonian} to the Hamiltonian~\eqref{eq:pspin-effective} means that the low part of the spectrum of $ \ham\ped{p}' $ has to match the spectrum of $ \ham\ped{p} $, and all other energy levels must be separated by a large energy gap from the original eigenvalues. Indeed, for the purpose of adiabatic quantum computation, only the ground state and the first excited subspace have to be matched in the purely adiabatic limit. However, in this paper we will always aim at matching the first $ L = 2^\nspin $ eigenvalues of $ \ham\ped{p}' $ and all the original spectrum. 
	We stress that even if the low-energy subspace of $ \ham\ped{p}' $ correctly reproduces the spectrum of $ \ham\ped{p} $, the quantum dynamics could be different. However, this mapping allows to solve the original optimization problem, through an experimentally viable effective model.
	
	Multiple-body interactions can be turned into $2$-body interactions using AND embedding. Pairs of binary variables $ (x_i, x_j) $ are encoded in an ancillary degree of freedom $ \tilde{x}_{i, j} = x_i \land x_j $. Of course, allowed configurations for the triple $ (x_i, x_j, \tilde{x}_{i, j}) $ are those where the logical AND is satisfied. A penalty function $ E\ped{pen}(x_i, x_j, \tilde{x}_{i, j}) $ penalizes nonphysical configurations through a large cost $ \delta > 0 $ or more. We will use the penalty function
	\begin{equation}\label{eq:penalty-function}
	    E\ped{pen}(x_i, x_j, \tilde{x}_{i,j}) = \delta(3 \tilde{x}_{i, j} + x_i x_j - 2 \tilde{x}_{i, j} x_i - 2 \tilde{x}_{i, j} x_j).
	\end{equation}
	It is easy to see that $ E\ped{pen} = 0 $ if $ \tilde{x}_{i, j} = x_i \land x_j $, while $ E\ped{pen} \ge \delta $ if $ \tilde{x}_{i, j} \ne x_i \land x_j $~\cite{biamonte:k-body-decomposition,zoller:decomposition}.
    
\begin{figure}[t]
        \centering
        \includegraphics[width = 0.7\textwidth]{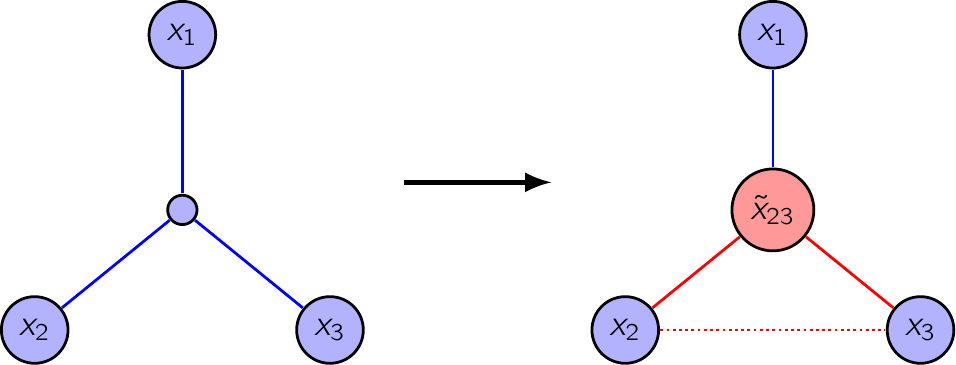}
        \caption{Visual representation of the AND embedding of Eq.~\eqref{eq:and-embedding}. On the left, the graph representing the original $3$-body term $J x_1 x_2 x_3$. On the right, the graph corresponding to the decomposed Hamiltonian with maximum $2$-body interactions, including penalties. Blue and red circles represent the original and the ancillary qubits, respectively. Blue lines represent coupling strength $J$, solid red lines represent $-2 \delta$ and dashed red lines $\delta$.}
        \label{fig:embedding-n-3}
    \end{figure}

    To be specific, consider a $3$-body term as $ J x_1 x_2 x_3$. Using the previously introduced AND embedding, this term can be rewritten using an ancillary qubit $\tilde{x}_{23}$ as
    \begin{equation}\label{eq:and-embedding}
        J x_1 x_2 x_3 \equiv J x_1 \tilde{x}_{23} + \delta(3\tilde{x}_{23} + x_2 x_3 - 2 \tilde{x}_{23} x_2 - 2\tilde{x}_{23} x_3),
    \end{equation}
    where the equivalence is intended as equality between corresponding $ L = 8 $ lowest energy levels. 
    This is pictorially represented in Fig.~\ref{fig:embedding-n-3}.
    
\section{A genetic algorithm for optimizing Hamiltonian free parameters}
\label{proposal}
This section is devoted to present the application of genetic algorithms for finding the free parameters useful for mapping $p$-body interacting systems in the $2$-body Hamiltonian. Genetic algorithms are population-based meta-heuristics which try to solve an optimization (or search)
problem by manipulating a multi-set of potential solutions and reproducing the natural selection process involving human individuals. In detail, as natural selection process leads to the survival of only the fittest human individuals (\ie, those capable of adapting to the changing environment), so the genetic algorithms perform an evolution process that leads to the survival of only the fittest solutions (\ie, those that better solve the optimization problem). Specifically, genetic algorithms operate on encoded representations of the solutions, called \emph{chromosomes}. To determine how good a solution is, a method named \emph{fitness function} is used to reflect the capability of the solution to solve the problem. In general, the workflow of a genetic algorithm includes the following steps. Firstly, a population of chromosomes is generated randomly and evaluated by using the fitness function. Successively, the algorithm performs a set of generations until some termination criteria are satisfied. In each generation, a set of chromosomes is selected to survive (\emph{parent selection mechanism}) and reproduce by means of the crossover operator. Generally, this operator takes in input two chromosomes (\emph{parent}) and gives in output two new chromosomes (\emph{offspring}) by exchanging portions of the parents. As in the natural evolution process, some mutations can occur. The mutation  operator performs by randomly changing some of the genes in the chromosomes. Both mutation and crossover operators are stochastic procedures that are applied according to a probability, named \emph{mutation probability} $p\ped{mut}$ and \emph{crossover probability} $p\ped{cx}$, respectively. As for the termination criteria, the most common one is the achievement of a maximum number of generations. Therefore, in this paper, we use this termination criterion.

Starting from this description, in order to implement a genetic algorithm for our problem, it is necessary to define the chromosome structure, the fitness function and the used genetic operators. Hereafter, a detailed description of the genetic algorithm components is given. 
3
\subsection{Chromosome structure}
The chromosome must encode the solution of our problem, that is the set of Hamiltonian free parameters~\eqref{eq:pspin-effective-classical}. In order to achieve this aim, the chromosome structure has been defined as follows:
\begin{equation}\label{eq:chromosome}
    \vec{v} \equiv (c_0, c_1, \dots, c_\ntot, d_{1, 2}, d_{1, 3}, \dots, d_{\ntot - 1, \ntot}).
\end{equation}
The length of the defined chromosome is $ D = (\ntot^2 + \ntot + 2) / 2 $. The values for the genes belong to the range $[-10, 10]$.  This choice is motivated by the fact that, in the analyzed cases, the genes of the chromosome $\vec{v}$ are strictly included within these bounds, except for the penalties that are not subject to the same constrictions.
    
\subsection{Fitness function}
The fitness function is used to evaluate the quality of the candidate solutions encoded in the chromosomes. It is implemented by taking into account $ E\ped{p} $ and $ E\ped{p}' $ reported in Eq.~\eqref{eq:pspin-hamiltonian-classical} and Eq.~\eqref{eq:pspin-effective-classical}, respectively. In detail, firstly, we list all possible configurations with $\nspin$ qubits, for the starting model, and with $ \ntot $ qubits, for the effective one. Conventionally, we arrange qubits of the effective model so that ancillae are at the beginning of the sequence. Secondly, we apply $ E\ped{p} $ and $ E\ped{p}' $ for each combination, sort the corresponding energies in ascending order and perform the differences.
    Formally, the fitness function $F$ is defined as follows:
     \begin{equation}\label{eq:fitness-function}
        F = \frac{1}{L}\sum_{i = 1}^{L} {\left[(E\ped{p})_{i} - (E\ped{p}')_{i}\right]}^2 + E\ped{pen}\api{vec},
    \end{equation}
    with $ L = 2^\nspin $. The first term enforces equality between corresponding eigenvalues, while the second one is a penalty cost to be applied when the eigenvectors of the effective Hamiltonian are ordered differently than the original ones. We do not apply penalties when eigenvectors are ordered differently within symmetry subspaces of the original Hamiltonian. In our code, $ E\ped{pen}\api{vec} = l \delta $, where $ l $ is the number of unsorted configurations.
	
\subsection{Genetic operators}
\label{operators}
Once defined the chromosome structure and the fitness function, it is necessary to discuss about the genetic operators, that is, crossover, mutation and selection mechanism. In the literature, different kinds of crossover, mutation and selection operators have been defined~\cite{YAO1993707,Herrera03ataxonomy}. However, when a new problem is addressed with genetic algorithms, it is necessary to select the most opportune configuration for these operators. For this reason, in this paper, we perform a design study of the implemented genetic algorithm aimed at selecting the most opportune configuration for the problem at issue. In detail, this study has involved the investigation of two different crossover operators, that is, the one-point crossover and the two point-crossover, different Gaussian distributions for mutation operator, and different values for tournament size for the selection mechanism. The results of this design study are reported in the next section. To conclude, in this section, we give more details about the investigated genetic operators. 
 \begin{description}        
 \item[Crossover operators] Generally, the crossover operator works by combining portions of two chromosomes, denoted as parents. In this work, we investigate two different strategies, \ie, one- and two-point crossover. In detail, the one point crossover chooses a random number $r$ in the range $[1, D-1]$ (with $D$ the length of the chromosome), and then splits both parents at this point by creating the two children by exchanging the tails. Instead, the two-point crossover chooses two random numbers $r_1$ and $r_2$ in the range $[1, D-1]$, breaks parents in these two points by creating the children by taking alternative
segments from the parents.
\item[Mutation] Generally, the mutation operator works by changing values of chromosome genes randomly. The Gaussian mutation chooses values drawn from a Gaussian distribution with zero mean and standard deviation $ \sigma$. In this work, we investigate several values for $ \sigma $, \ie, $ \sigma = \text{\numlist{0.2;0.4;0.6;0.8;1.0}} $. 
 \item[Selection] Selection mechanism is devoted to select the chromosomes that will become parents of the next generation. One of the most known selection operators is the tournament mechanism which selects each parent by performing a tournament among $\tournsize$ chromosomes, randomly selected, where the chromosome that wins is the fittest one. In this work, we investigate $ \tournsize = \text{\numlist{2;3;5}} $.
    \end{description}

\section{Preliminary experiments and results}
\label{sec:Experimental Results}
This section is devoted to show the results of some preliminary experiments carried out to demonstrate the feasibility of the proposed approach. In detail, the designed genetic algorithm is applied to solve two simple configurations of the ferromagnetic $p$-spin model. This choice is due to the possibility to analytically solve these configurations and perform a comparison with the output of the genetic algorithm. The configuration of the applied genetic algorithm is the result of a design study involving the genetic operators described in Section~\ref{operators}. The comparison between the solution obtained by the designed genetic algorithm and that computed analytically is carried out by considering the energy eigenvectors and eigenvalues of the first $2^\nspin$ Hamiltonian states, as well as the Hamiltonian free parameters. Moreover, the use of the solution obtained by the genetic algorithm is investigated for the adiabatic quantum computation with respect to the original $p$-spin model. Hereafter, more details about the considered configurations of the ferromagnetic $p$-spin model, the design study, the comparison results and the exploitation of genetic solutions in the adiabatic quantum computation are given.

\subsection{Experimental set-up}
\label{setup}
\begin{figure}
    \centering
    \includegraphics[width = 0.35\textwidth]{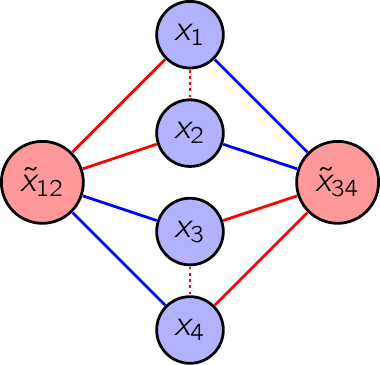}
    \caption{Graph representing the effective $2$-body model for the $ p $-spin Hamiltonian with $\nspin = 4$ and $ p = 3 $, originally containing four $3$-body terms. Blue and red circles represent the original and the ancillary qubits, respectively. Blue lines represent coupling strength $J$, solid red lines represent $-2 \delta$ and dashed red lines $\delta$.}
    \label{fig:and-embedding-n-4}
\end{figure}

To perform our experimentation, we consider the simplest configurations of the ferromagnetic $ p $-spin model which require the minimum number of ancillary qubits for embedding, \eg, $\nspin=3,\, p=3$ and $\nspin=4,\, p=3$. The Hamiltonian of the former one only contains a single $3$-body term, which can be decomposed as described in Eq.~\eqref{eq:and-embedding} with a single ancilla, \ie,  $\tilde{x}_{23} = x_2 \land x_3$ and hence $\ntot = 4$. By contrast, the Hamiltonian for the $ \nspin = 4$ case contains four $3$-body terms, which require two ancillae, \ie, $\tilde{x}_{12} = x_1 \land x_2$ and $\tilde{x}_{34} = x_3 \land x_4$, to be decomposed as described in Section~\ref{sec:problem}, \ie, $ \ntot = 6 $. The reduction process leads to the graph represented in Fig.~\ref{fig:and-embedding-n-4}.

These two cases are selected because it is possible to work out by hand the analytic solution for these settings with little effort and, this is useful for carrying out the comparison study with the designed genetic algorithm. We report the analytic solutions below. 
\begin{gather}
    \vec{v}_{\ntot=4} \equiv (-3,\, -3\delta,\, \frac{26}{9},\, \frac{26}{9},\, \frac{26}{9},\, 2\delta,\, 2\delta,\, \frac{16}{3},\, -\frac{8}{3}-\delta,\, -\frac{8}{3},\, -\frac{8}{3});\label{eq:analytic-chromosome}\\[1ex]
\begin{align}
  \vec{v}_{\ntot=6} \equiv {}&(-4,\,-3\delta,\,-3\delta,\,\frac{7}{2},\,\frac{7}{2},\,\frac{7}{2},\,\frac{7}{2},\,0,\,2\delta,\,2\delta,\notag\\
&3,\,3,\,3,\,3,\,2\delta,\,2\delta,\,-3-\delta,\,-3,\,-3,\,-3,\,-3,\,-3-\delta).
\end{align}
\end{gather}
In what follows, we will fix $ \delta = 50 $ as this number provides a large separation between the largest eigenvalue of the target subspace and the smallest eigenvalue of the nonphysical one, in both cases.

\subsection{Design study}
\label{design}
In order to select the best configuration for genetic operators, we perform a design study by considering the operators described in Section~\ref{operators}. By using $2$ different crossover operators, $5$ different mutation operators and $3$ different selection operators, our design study involves the assessment of $30$ different combinations. Table~\ref{tab:1} gives an index to the different combinations. As for the other parameters of the genetic algorithm, in our experimentation, we set $\npop = 20$ chromosomes, the crossover probability $ p\ped{cx} = 0.4 $, the mutation probability $ p\ped{mut} = 0.7 $. This choice is not typical, as usually $ p\ped{mut} < p\ped{cx} $. However, the results we discuss below are qualitatively independent on these two parameters. The termination criterion is the achievement of a number of generations, \ie, $\ngen = \num{25000}$.
Genetic algorithms are stochastic procedures, thus we repeat the simulation $ 100 $ times for every combinations. 

\begin{figure}
    \centering
    \includegraphics[width = 0.9\textwidth]{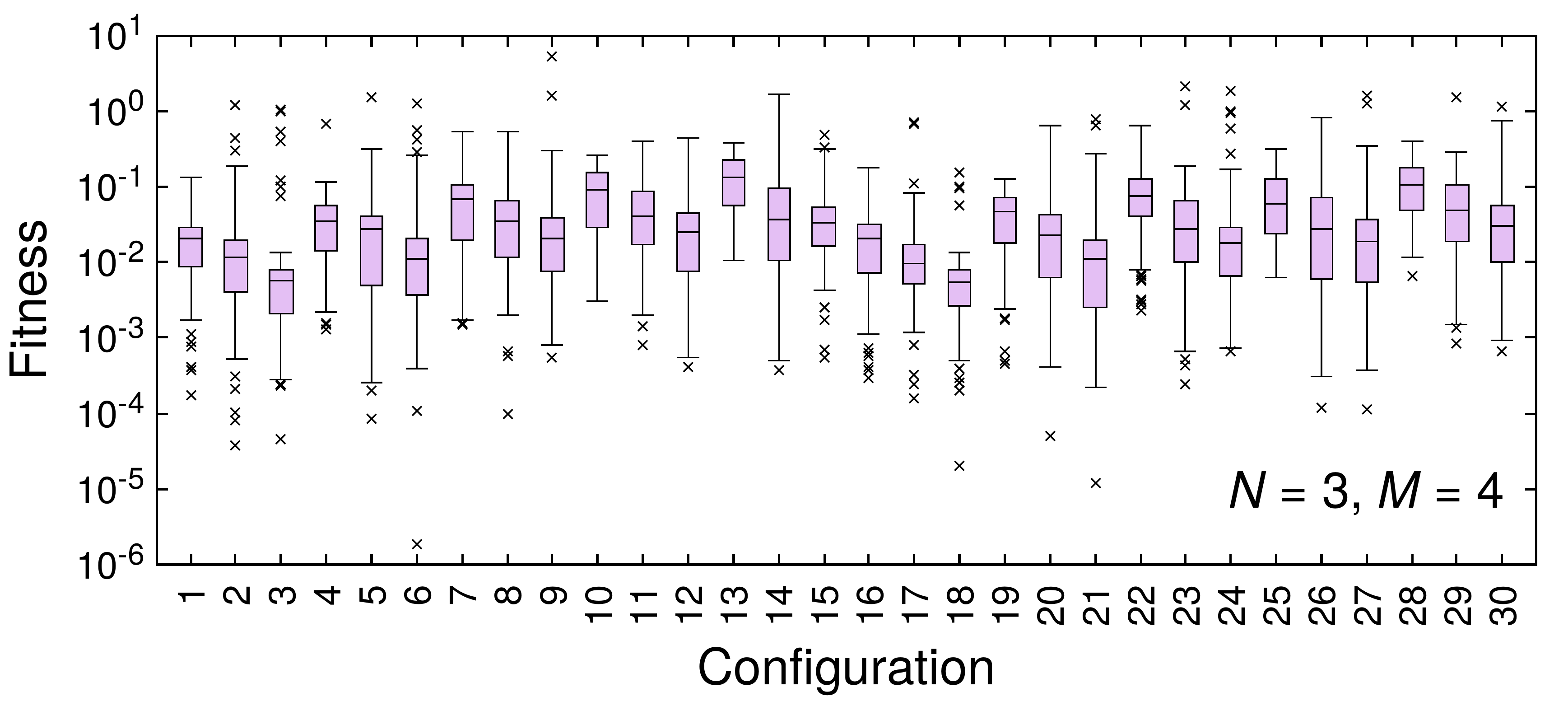}\\\includegraphics[width = 0.9\textwidth]{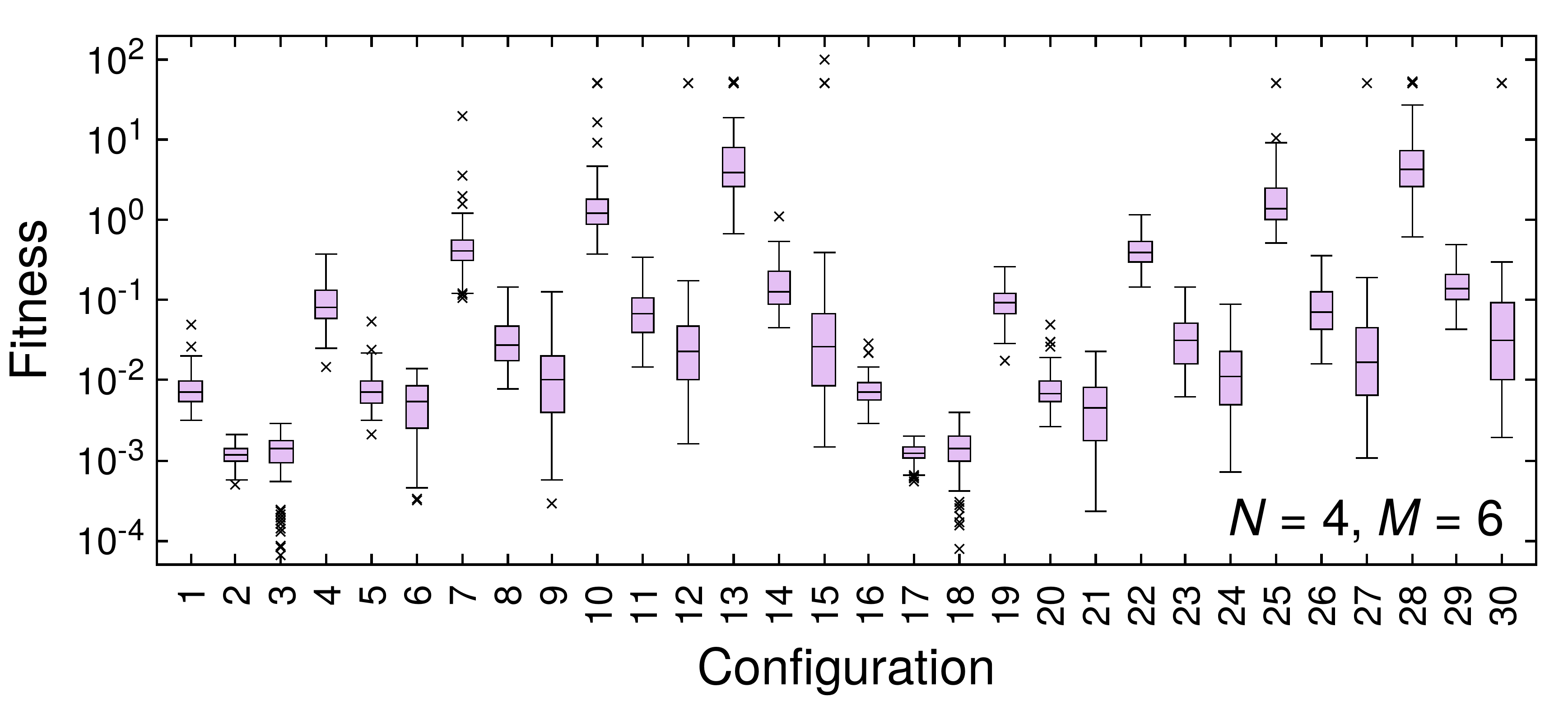}
    \caption{Box graphs depicting the distribution of fitness values for different combinations of the genetic operators. For each combination, the black line inside the box corresponds to the median over $ 100 $ runs of the genetic algorithm. Outliers are explicitly indicated using black crosses. The indexes of the combinations on the $ x $-axis are tabulated in Table ~\ref{tab:1}.}
    \label{fig:parameter_configuration}
\end{figure}

The comparison among all the different combinations of genetic operators is shown in the boxplot of Fig.~\ref{fig:parameter_configuration}. In detail, boxplots show the minimum, the maximum, the median and the likely range of variation of the fitness values over the $100$ runs. However, in order to select the most opportune combination, the median fitness values are compared.

By analyzing Fig.~\ref{fig:parameter_configuration}, for $\nspin = 3$, $\ntot = 4$, the best median of the fitness values (the minimum one) is the combination $18$, \ie, the combination involving the two-point crossover, the Gaussian mutation with $\sigma = 0.2$ and tournament selection with $\tournsize = 5$. Instead, for $ \nspin = 4 $ and $ \ntot = 6$, the configuration $ 2 $ is the one yielding the smallest median fitness value, \ie, the combination involving one-point crossover, $\sigma = 0.2$ and $\tournsize = 3$.

\begin{table}
	\centering
	\caption{Combinations of genetic operators investigated in the design study. 1P (2P) stands for one-point (two-point) crossover.}
	\label{tab:1}
	\begin{tabular}{rccc@{\hskip 0.75in}rccc}
		\toprule
		\# & Crossover & $\sigma$  & $ N_\text{T} $ & \# & Crossover & $\sigma$  & $ N_\text{T} $  \\
		\midrule
		$1$  & 1P & $0.2$ & $2$ & $16$ & 2P & $0.2$ & $2$ \\   
		$2$  & 1P & $0.2$ & $3$ & $17$ & 2P & $0.2$ & $3$ \\   
		$3$  & 1P & $0.2$ & $5$ & $18$ & 2P & $0.2$ & $5$ \\  
		$4$  & 1P & $0.4$ & $2$ & $19$ & 2P & $0.4$ & $2$ \\   
		$5$  & 1P & $0.4$ & $3$ & $20$ & 2P & $0.4$ & $3$ \\ 
		$6$  & 1P & $0.4$ & $5$ & $21$ & 2P & $0.4$ & $5$ \\ 
		$7$  & 1P & $0.6$ & $2$ & $22$ & 2P & $0.6$ & $2$ \\  
		$8$  & 1P & $0.6$ & $3$ & $23$ & 2P & $0.6$ & $3$ \\ 
		$9$  & 1P & $0.6$ & $5$ & $24$ & 2P & $0.6$ & $5$ \\ 
		$10$ & 1P & $0.8$ & $2$ & $25$ & 2P & $0.8$ & $2$ \\   
		$11$ & 1P & $0.8$ & $3$ & $26$ & 2P & $0.8$ & $3$ \\ 
		$12$ & 1P & $0.8$ & $5$ & $27$ & 2P & $0.8$ & $5$ \\ 
		$13$ & 1P & $1.0$ & $2$ & $28$ & 2P & $1.0$ & $2$ \\ 
		$14$ & 1P & $1.0$ & $3$ & $29$ & 2P & $1.0$ & $3$ \\
		$15$ & 1P & $1.0$ & $5$ & $30$ & 2P & $1.0$ & $5$ \\
		\bottomrule
	\end{tabular}
\end{table}

\subsection{Results}
\label{results}
\begin{figure}
    \centering
    \includegraphics[width = 0.6\textwidth]{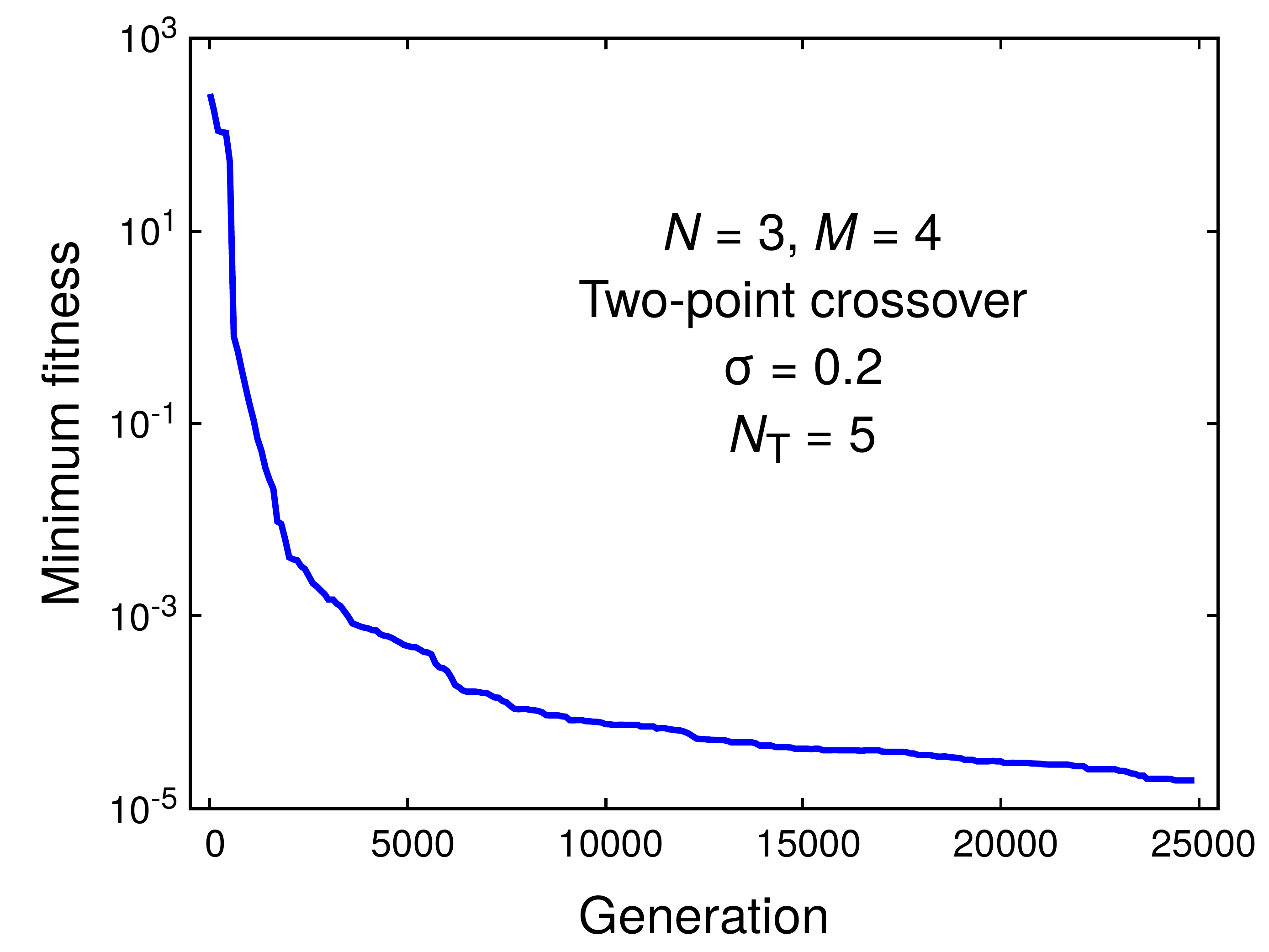}
    \caption{Updating of the fitness values through \num{25,000} generations, for the case $ \nspin = 3$,  $\ntot = 4$, and for the best combination of genetic operators, \ie, combination $18$ in Table ~\ref{tab:1}. The final fitness value at the last generation is $ F = \num{9.8799e-8} $.}
\label{fig:convergence}       
\end{figure}

Once performed $100$ runs of the genetic algorithm with configuration $18$ for $ \nspin = 3 $, $ \ntot = 4 $ and $100$ runs with configuration $2$ for $ \nspin = 4 $, $ \ntot = 6 $, we obtain $100$ solutions for $ \nspin = 3 $, $ \ntot = 4 $ and $100$ solutions for $ \nspin = 4 $, $ \ntot = 6 $. As an example, Fig.~\ref{fig:convergence} shows the fitness values against the number of generations for the genetic algorithm with configuration $18$ used to address $ \nspin = 3 $, $ \ntot = 4 $ problem.


To compare the solutions obtained by the genetic algorithm and the analytically computed ones for both considered configurations of the ferromagnetic $p$-spin model, we select the best chromosome (\ie, the one with the least fitness value) among all solutions over $100$ runs. The comparison is carried out by considering the computed Hamiltonian free parameters, but also the energy eigenvectors and eigenvalues of the first $2^\nspin$ states of the embedded Hamiltonian generated by the genetic algorithm with that of the original Hamiltonian in Eq.~\eqref{eq:pspin-hamiltonian-classical}. Table~\ref{tab:2} and Table~\ref{tab:3} show the results of this comparison for $\nspin = 3$, $\ntot = 4$ and $\nspin = 4$, $\ntot = 6$, respectively. For $ \nspin = 3 $, $ \ntot = 4 $, the first qubit in the genetic eigenvector is the ancilla qubit defined as $ q_0 = q_1 \land q_2 $. For the case $ \nspin = 4 $ and $ \ntot = 6 $, the first two qubits of the sequence are the two ancillae, defined as $ q_0 = q_2 \land q_3 $ and $ q_1 = q_4 \land q_5 $.  In addition, the first $2^{\nspin}$ genetic eigenvectors always respect the original degeneracies of the starting spectrum. Moreover, we also observe the sign-flip pattern in the spectrum, as predicted by the $Z_2$ anti-symmetry of this model for odd $ p $. 

An indicator of the accuracy of the returned solution is the root mean square
\begin{equation}\label{eq:rms}
    \text{rms} \equiv \sqrt{\frac{1}{D} \sum_{i = 1}^{D} {\left(\frac{v_i\api{analytic} - v_i\api{genetic}}{v_i\api{analytic}}\right)}^2}.
\end{equation}
In the $ \nspin = 3 $ case, the best solution yields $\text{rms} \approx \num{7.2e-3}$, while for $ \nspin = 4 $ we have $ \text{rms} \approx 0.22 $. The scaling of the accuracy of the best returned solution as a function of the input size is a serious question that will be addressed in a forthcoming paper. However, we observe that the analytic solution is qualitatively well-reproduced by the genetic algorithm in both cases.



\begin{table}[tb]
	\centering
	\caption{The results of the comparison between the best chromosome obtained by the genetic algorithm and the analytically computed solution for $\nspin = 3$, $\ntot = 4$ problem. We fixed $\delta = 50$.}
	\label{tab:2}
	\begin{tabular}{cccccc}
		\toprule
		\multicolumn{2}{c}{Free parameters} & \multicolumn{2}{c}{Eigenvectors} & \multicolumn{2}{c}{Eigenvalues}\\
		Analytic   &  {Genetic} & Analytic   &  Genetic & {Analytic} & {Genetic} \\
		\midrule
		$-3        $   & $ -2.99919 $ & [0, 0, 0] & [0, 0, 0, 0] & $ -3.00000 $   & $ -2.99919 $ \\    
		$-150      $   & $ -150.853 $ & [0, 0, 1] & [0, 1, 0, 0] & $ -0.11111 $   & $ -0.11138 $ \\   
		$ 26 / 9   $   & $ 2.88781  $ & [0, 1, 0] & [0, 0, 0, 1] & $ -0.11111 $   & $ -0.11129 $ \\    
		$ 26 / 9   $   & $ 2.88795  $ & [1, 0, 0] & [0, 0, 1, 0] & $ -0.11111 $   & $ -0.11124 $ \\    
		$ 26 / 9   $   & $ 2.88790  $ & [0, 1, 1] & [1, 1, 1, 0] & $ 0.11111  $   & $ 0.11111  $ \\    
		$ 100      $   & $ 100.720  $ & [1, 0, 1] & [0, 1, 0, 1] & $ 0.11111  $   & $ 0.11120  $ \\    
		$ 100      $   & $ 101.118  $ & [1, 1, 0] & [0, 0, 1, 1] & $ 0.11111  $   & $ 0.11120  $ \\    
		$ 16 / 3   $   & $ 5.33174  $ & [1, 1, 1] & [1, 1, 1, 1] & $ 3.00000  $   & $ 2.99999  $ \\    
		$ -158 / 3 $   & $ -53.6496 $   \\  
		$ -8 / 3   $   & $ -2.66531 $   \\ 
		$ -8 / 3   $   & $ -2.66545 $   \\
		\bottomrule
	\end{tabular}
\end{table}

\begin{table}[tb]
	\centering
	\caption{The results of the comparison between the best chromosome obtained by the genetic algorithm and the analytically computed solution for $\nspin = 4$, $\ntot = 6$ problem. We fixed $\delta = 50$.}
	\label{tab:3}
	\begin{tabular}{cccccc}
		\toprule
		\multicolumn{2}{c}{Free parameters} & \multicolumn{2}{c}{Eigenvectors} & \multicolumn{2}{c}{Eigenvalues}\\
		Analytic &  Genetic & Analytic &  Genetic & Analytic &  Genetic \\
		\midrule
		 $-4$    & $-3.99450$   & [0, 0, 0, 0] & [0, 0, 0, 0, 0, 0] & $-4.0$ & $-3.99450$ \\
		 $-150$  & $-148.165$  & [0, 0, 0, 1] & [0, 0, 0, 0, 1, 0] & $-0.5$ & $-0.53947$ \\	
		 $-150$  & $-144.833$ & [0, 0, 1, 0] & [0, 0, 1, 0, 0, 0] & $-0.5$ & $-0.52837$ \\	
		 $7 / 2$ & $3.46613$    & [0, 1, 0, 0] & [0, 0, 0, 0, 0, 1] & $-0.5$ & $-0.47565$ \\
		 $7 / 2$ & $3.54304$    & [1, 0, 0, 0] & [0, 0, 0, 1, 0, 0] & $-0.5$ & $-0.45146$ \\
		 $7 / 2$ & $3.45503$	& [0, 0, 1, 1] & [1, 0, 1, 1, 0, 0] & $-0.0$ & $-0.03157$ \\
		 $7 / 2$ & $3.51886$    & [0, 1, 0, 1] & [0, 0, 0, 1, 0, 1] & $-0.0$ & $-0.02561$ \\
		 $0$     & $-0.02015$   & [0, 1, 1, 0] & [0, 0, 1, 0, 0, 1] & $-0.0$ & $-0.01985$ \\
		 $100$   & $96.397$	& [1, 0, 0, 1] & [0, 0, 1, 0, 1, 0] & $-0.0$ & $-0.01965$ \\	
		 $100$   & $95.7088$   & [1, 0, 1, 0] & [0, 0, 0, 1, 1, 0] & $-0.0$ & $0.00748$  \\
		 $3$     & $2.98789$    & [1, 1, 0, 0] & [0, 1, 0, 0, 1, 1] & $-0.0$ & $0.04105$  \\
		 $3$     & $3.12228$    & [0, 1, 1, 1] & [1, 0, 1, 1, 1, 0] & $0.5$  & $0.46894$  \\
		 $3$     & $2.91879$    & [1, 0, 1, 1] & [0, 1, 1, 0, 1, 1] & $0.5$  & $0.46932$  \\
		 $3$     & $3.04070$    & [1, 1, 0, 1] & [1, 0, 1, 1, 0, 1] & $0.5$  & $0.50621$  \\	
		 $100$   & $97.8836$	& [1, 1, 1, 0] & [0, 1, 0, 1, 1, 1] & $0.5$  & $0.53568$  \\
		 $100$   & $97.9453$   & [1, 1, 1, 1] & [1, 1, 1, 1, 1, 1] & $4.0$  & $4.00773$  \\	
		 $-53$   & $-58.5698$  \\
		 $-3$    & $-2.94631$   \\
		 $-3$    & $-3.01034$   \\
		 $-3$    & $-2.99610$   \\	
		 $-3$    & $-3.09301$   \\
		 $-53$   & $-56.9343$  \\
		\bottomrule
	\end{tabular}
\end{table}

\subsection{Discussion for adiabatic quantum computation}

The genetic $2$-body model can be used for adiabatic quantum computation, with the time-dependent Hamiltonian of Eq.~\eqref{eq:time-dependent-hamiltonian}, and compared with the original $p$-spin model, or with the analytic $2$-body model. In this last part, we focus on $ \nspin = 3 $, $ \ntot = 4 $ for computational ease. We performed the same analysis also for $ \nspin = 4 $, $ \ntot = 6 $ with similar results. For the purpose of quantum optimization, it is paramount that the fidelity $\Phi$, \ie, the ground state occupation probability at the end of the annealing ($ s = 1 $), is large. Of course, due to the larger number of degrees of freedom of the effective model with ancillae, we expect that a slower annealing is needed to reach the target ground state, compared with the original $ p $-spin model.

\begin{figure}[htb]
\centering
 \subfloat[]{{\includegraphics[width = 0.45\textwidth]{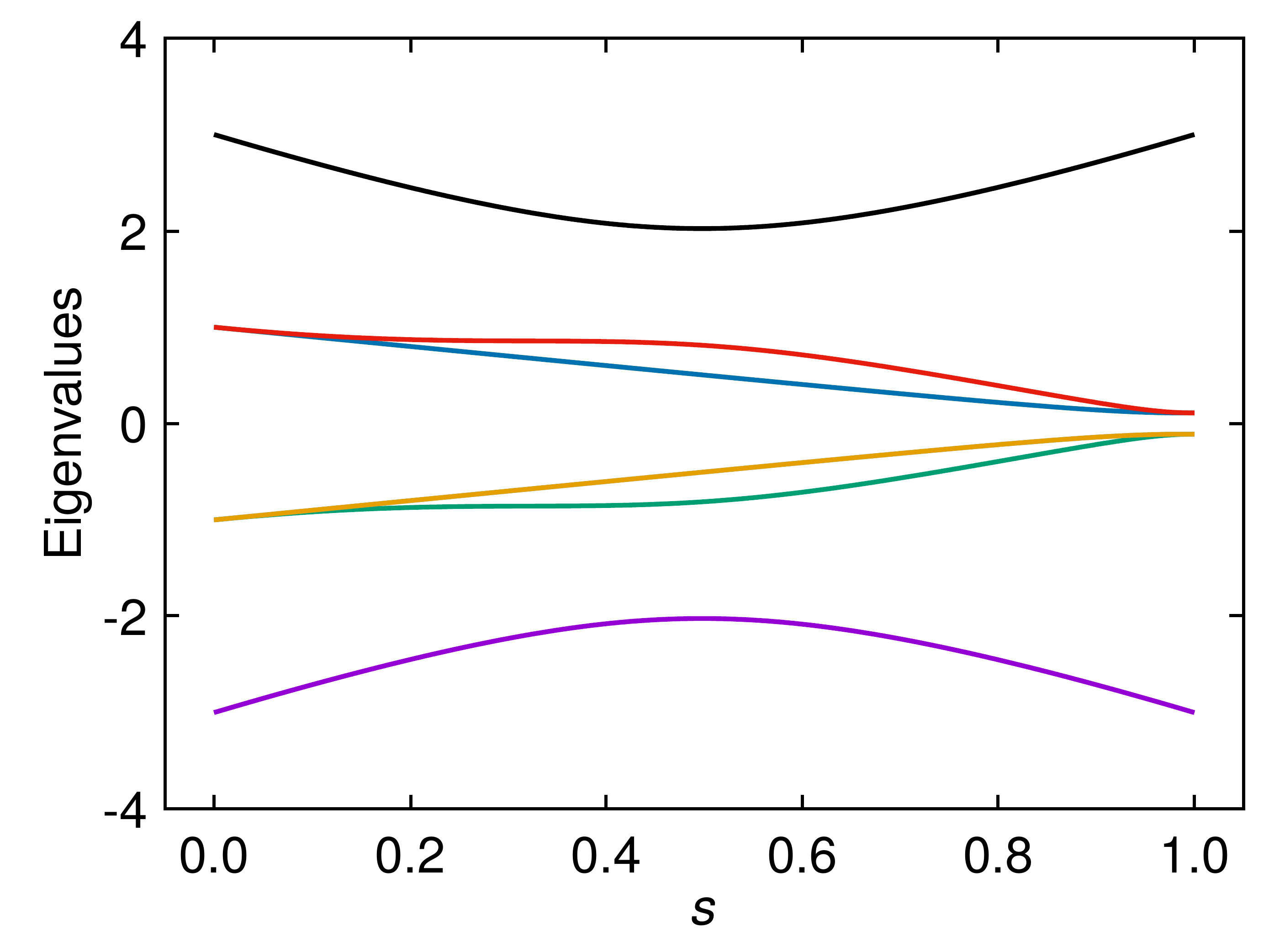}}}%
\qquad
\subfloat[]{{\includegraphics[width = 0.45\textwidth]{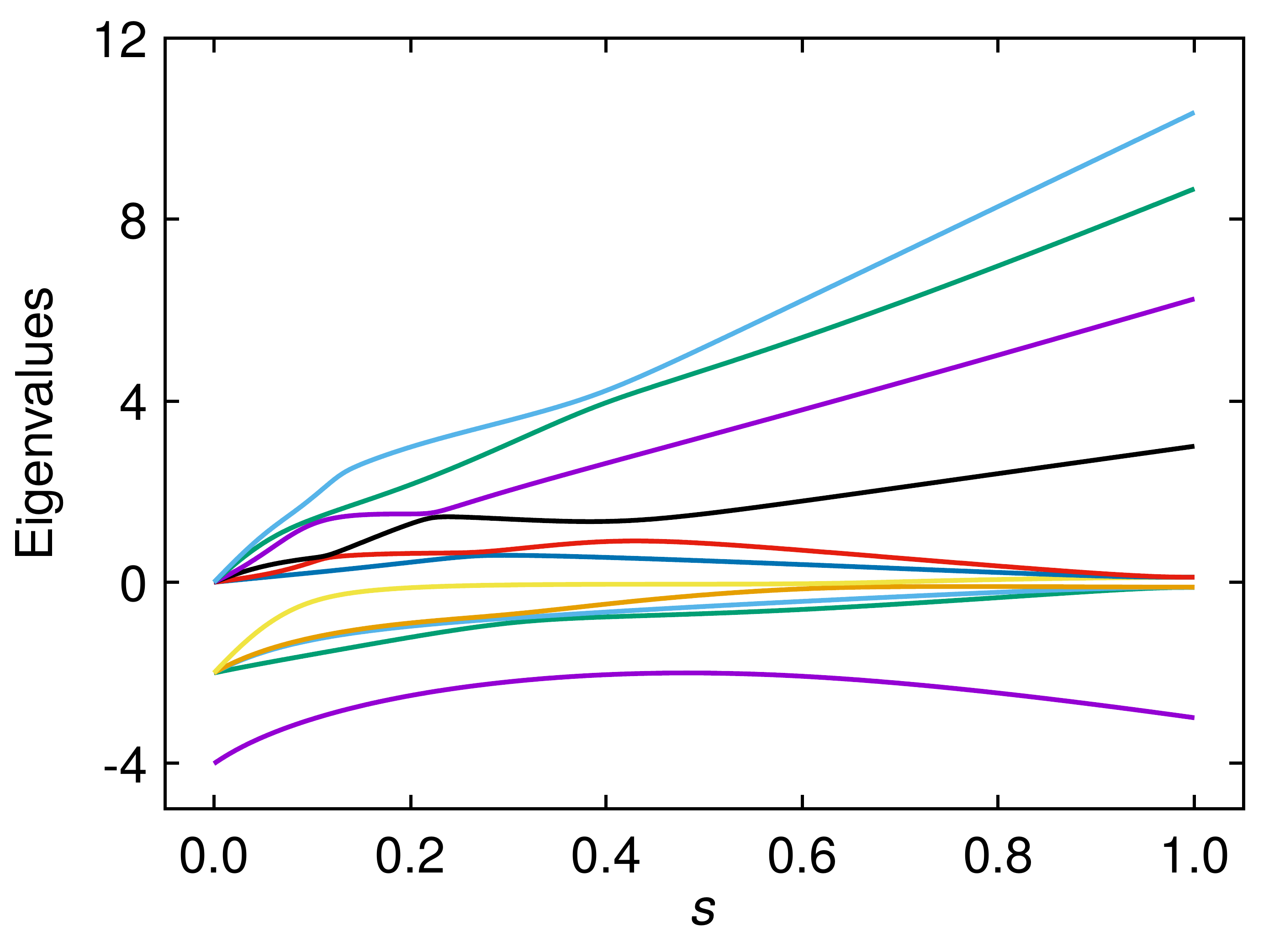}}}%
\caption{Instantaneous eigenvalues of the time-dependent Hamiltonian~\eqref{eq:time-dependent-hamiltonian}, for $ s \in \left[0, 1\right] $, with $ \delta = 11 $ for sake of clarity. Panel~(a) is for the $p$-body ferromagnetic $p$-spin model in Eq.~\eqref{eq:pspin-hamiltonian}, panel~(b) is for the genetic $2$-body Hamiltonian in Eq.~\eqref{eq:pspin-effective}, for the first $ L = 11 $ states. At $ s = 1 $, the first $2^\nspin$ states are the same for the two models, whereas higher (penalized) energy levels are separated from the low part of the spectrum. Increasing $\delta$ will increase this separation.}%
\label{fig:Annealing_dynamics}
\end{figure}

First, we compare the low part of the instantaneous spectra of the two models in Fig.~\ref{fig:Annealing_dynamics}, using $ \delta = 11 $ for visual clarity. We observe that the first $2^\nspin = 8$ states match at $ s = 1 $, though they differ for $ 0 < s < 1  $. Higher excited states, subjected to penalty, are significantly separated from the lower ones.

\begin{figure}[tb]
    \centering
    \includegraphics[width = 0.6\textwidth]{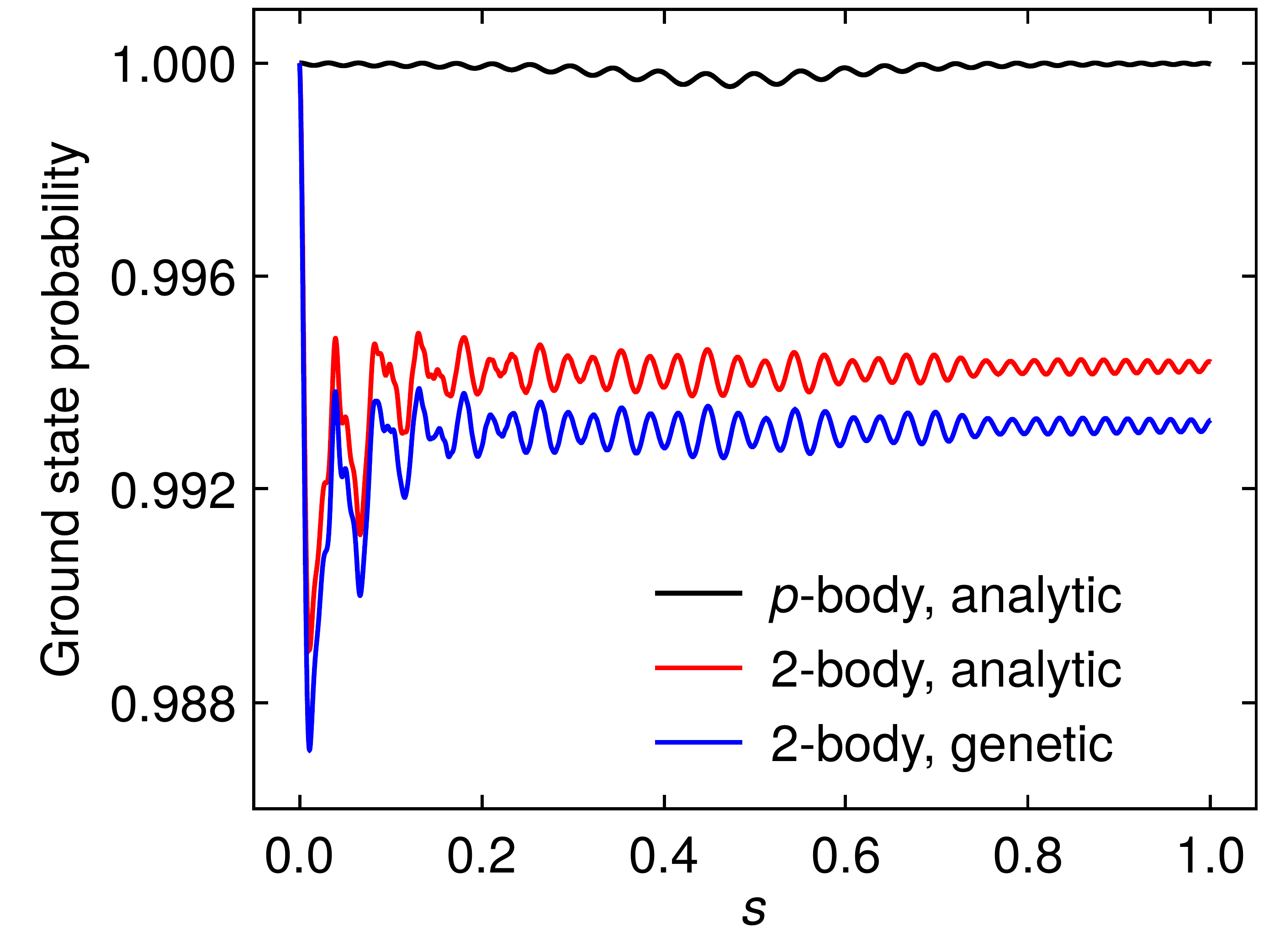}
    \caption{Evolution of the ground state occupation probability as a function of the dimensionless time $ s $, for the original $ p $-spin model with $ \nspin = 3 $ (black line), the analytic $ 2 $-body model of Eq.~\eqref{eq:analytic-chromosome} (red line), and the best genetic solution (blue line).}
    \label{fig:annealing}
\end{figure}

Second, we classically simulate a quantum evolution of annealing time $ \tf = 100 $, and compare the evolution of the ground state occupation probability of the genetic model with those of the original $ p $-spin model and of the analytic $2$-body model of Eq.~\eqref{eq:analytic-chromosome}. Results are shown in Fig.~\ref{fig:annealing}. The ground state population of the effective model evolves differently that the original one. This is not surprising, as the goal of our genetic algorithm is to match the final spectrum, irrespective of the instantaneous dynamics. By contrast, the evolution of the genetic model closely resembles that of the analytic $2$-body model. The fidelity at the end of the evolution is large ($ \Phi \approx 0.994 $ and $ \Phi \approx 0.993 $ for the analytic and the genetic $2$-body Hamiltonians, respectively), although not as large as that of the original model ($ \Phi \approx 0.99998 $) for this choice of $ \tf $. This can be justified by the adiabatic condition. 

In fact, a common adiabatic criterion states that the evolution time must satisfy the following condition,
\begin{equation}\label{eq:adiabatic}
    \tf \gg \max_{a, b, s} \frac{\braket{\epsilon_a(s) | \ham(s) | \epsilon_b(s)}}{\gap^2},
\end{equation}
where $ \ket{\epsilon_a(s)} $ are the instantaneous eigenstates of $ \ham(s) $~\cite{albash:review-aqc}. The introduction of ancillary qubits with large energy penalties $\delta$ for unphysical configuration makes the numerator of the right-hand side of Eq.~\eqref{eq:adiabatic} larger for the effective model than for the original $p$-spin model, while the minimal gap is similar for both models. For the cases we analyzed, the adiabatic time scale of the effective model is $\sim \delta$ times longer than for the original model. Thus, it is natural to expect that, for fixed $ \tf $, the original model is closer to the adiabatic limit than the effective one, thus the corresponding fidelity is larger.

\section{Conclusions and future research directions}
\label{sec:Conclusions}

Using a genetic algorithm, we have mapped the ferromagnetic $p$-spin Hamiltonian into a Hamiltonian with only $2$-body interactions. We have shown, in two analytically solvable cases, that our strategy can successfully be used for this task. In fact, the energy eigenvalues and eigenvectors of the first $2^\nspin$ states of the original Hamiltonian are correctly reproduced, with rms (Eq.~\eqref{eq:rms}) of $\approx 1.19\times 10^{-3}$ for the best combination of genetic operators for the case of $\ntot=4$. 
However, the considered configurations of the ferromagnetic $p$-spin model are the simplest ones. In the future, a wider experimentation involving higher configurations (\ie, larger integral values for $\nspin$ and $p$) will be carried out to show the benefits of our proposal. Since higher configurations represent harder problem instances, dealing with them could require to change genetic algorithm parameters by increasing, for example, the population size or the number of maximum generations. Moreover, the complexity of dealing with higher configurations could open the doors to the application of new evolutionary algorithms such as memetic algorithms~\cite{moscato1989evolution}, \ie, population-based meta-heuristics combining global search with local search procedures. Finally, for larger systems, the required number of ancillae becomes non-trivial and, as a consequence, improvements should be done to address also this problem. We could also use our technique to predict the minimum number of ancillae required for the embedding, in the case of large systems where the analytical mapping could be cumbersome.



\end{document}